\pdfoutput=1
\documentclass[fullpage]{article}
\usepackage{graphicx}
\usepackage{amsfonts}
\usepackage{xcolor}
\usepackage{tikz}
\usepackage{slashbox}
\usepackage{varwidth}
\usepackage{caption}
\usepackage{subcaption}
\usepackage{authblk}
\usepackage{booktabs}
\usepackage{multirow}
\usepackage{float}
\usepackage{amsthm}
\usepackage{amsmath}
\usepackage{bbm}
\usepackage{mathrsfs}
\usepackage{bm}
\usepackage{url}
\usepackage{makecell}

\renewcommand\theadfont{\bfseries}

    \usepackage[T1]{fontenc}
    \usepackage{mathpazo}
    \usepackage{adjustbox} 
    \usepackage{xcolor} 
    \usepackage{enumerate} 
    \usepackage{geometry} 
    \usepackage{amsmath} 
    \usepackage{amssymb} 
    \usepackage{textcomp} 
    \AtBeginDocument{%
    }
    \usepackage{upquote} 
    \usepackage{eurosym} 
    \usepackage[mathletters]{ucs} 
    \usepackage[utf8x]{inputenc} 
    \usepackage{fancyvrb} 
    \usepackage{grffile} 
    \usepackage{hyperref}
    \usepackage{longtable} 
    \usepackage{booktabs}  
    \usepackage[inline]{enumitem} 
    \usepackage[normalem]{ulem} 

\DeclareMathOperator*{\argmin}{arg\,min}

\usetikzlibrary{arrows}
\usetikzlibrary{shapes}
\title{Estimating Future VaR from Value Samples and Applications to Future Initial Margin}
\date{}
\author{Narayan Ganesan, Bernhard Hientzsch}
\affil{Corporate Model Risk, Wells Fargo}

\begin{document}
\maketitle

\begin{abstract}
Predicting future values at risk (fVaR) is an important problem in finance. They
arise in the modelling of future initial margin requirements for counterparty
credit risk and future market risk VaR. One is also interested in derived
quantities such as: i) Dynamic Initial Margin (DIM) and Margin Value Adjustment
(MVA) in the counterparty risk context; and ii) risk weighted assets (RWA) and
Capital Value Adjustment (KVA) for market risk. This paper describes several
methods that can be used to predict fVaRs. We begin with the Nested MC-empirical
quantile method as benchmark, but it is too computationally intensive for
routine use. We review several known methods and discuss their novel applications
to the problem at hand. 
The techniques considered include computing percentiles from distributions 
(Normal and Johnson) that were matched to parametric moments or percentile
estimates, quantile regressions methods, and others with more specific
assumptions or requirements. 
 We also consider how limited inner simulations can be used to
improve the performance of these techniques. The paper also provides illustrations, results,
and visualizations of intermediate and final results for the various approaches
and methods.

\end{abstract}

\section{Introduction}

\subsection{Portfolios, Cashflows, and
Values}\label{portfolios-cashflows-and-values}

Consider a generic financial portfolio described by a complete set of
cashflows \({\bf CF}=(t_i,CF_i)_{i \in {\cal I}}\) where \({\cal I}\) is an appropriate
index set. Each cashflow \(CF_i\) is determined from values of ''underliers''
\({\bf X}_s\) for \(s\leq t\), where the underliers \({\bf X}_t\) represent a
certain stochastic process. Formally \(CF_i\) is measurable as of time \(t_i\)
with respect to a certain filtration \({\cal F}_t\) which is generated by \({\bf
X}_s\) for \(s \leq t\). Examples of the financial could range from a single
European option on a single underlier to a large portfolio representing a
complete netting set of many financial instruments depending on many underliers.

In addition, we suppose that the value of the outstanding cashflows of the
portfolio at time \(t\) (in an appropriate sense such as replication by the bank
or a computation by some accepted central calculation agent) is given as: i) a
stochastic process \(V_t\) or ii) as an exact or approximate function \(V(t,{\bf
X}_t)\) (or \(V(t,{\bf X}_t,{\bf U}_t)\) where \({\bf U}_t\) is extra Markovian
state necessary to compute the value, for instance a barrier breach indicator
for barrier options).
For approximate computation, we can view it as \(V(t,{\bf R}_t)\) where \({\bf
R}_t\) are an appropriate set of features measurable as of \({\cal F}_t\). We
also assume that the \(V_t\)'s for different \(t\) are discounted to a common
time or otherwise made comparable so that they can be added and subtracted to or
from each other and any \(CF_i\) meaningfully.

We can interpret \(V_t\) as the value to the bank ("us"); positive \(CF_i\)
would be interpreted as payments to us, while negative \(CF_i\) are interpreted
as payments by us to some counterparty.

Assuming some realization of \({\bf X}_t\) under some measure \({\cal M}\),
indicated by an argument \(\omega\) (as in \({\bf X}_t(\omega)\)), there is a
corresponding realization \(V_t(\omega)\) and \({\bf CF}(\omega)\). 
When comparing between different measures, if \(V\) is given by a function, 
it will be the same function but its arguments will follow their dynamics
under the measure \({\cal M}\), while the processes \(V_t\) will be different.
In our examples here, the \({\bf X}_t\) will be  given as realizations of a system of SDEs.

\subsection{\texorpdfstring{Change of Value over a Time Period \(t\)
to
\(t+\delta\)}{Change of Value over a time period t to t+\textbackslash{}delta}}\label{change-of-value-over-a-time-period-t-to-tdelta}

Consider the change of value over the time period \(t\) to
\(t+\delta\) defined by
\begin{equation}
\Delta V^{\delta,f,{\cal M}}_t (\omega) := V_{t+\delta} (\omega) + \sum_{i:t_i \in [t,t+\delta) } f(CF_i(\omega),t_i(\omega),t) - V_t (\omega)
\end{equation}
If the measure ${\cal M}$ or function \(f\) are clear from the context, we omit the
corresponding superscripts.

Some simple examples for the function \(f(x,t_i,t)\) are
\(x\), \(0\), \(x^{+}\), or \(x^{-}\). The interpretation would be that cash
flows from the portfolio during that period are assumed to be not made or only
partially made. In the counterparty credit risk applications, one
could assume that the period from \(t\) to \(t+\delta\) - the ``margin period
of risk" (MPoR) - describes the period
from the first uncertainty about the default of the counterparty to its
resolution by default and liquidation and/or replacement of the portfolio.
Often one assumes that the bank will make payments until default is
almost certain while the counterparty would avoid payments,
for instance by engaging in disputes. For market risk purposes, all
payments are typically assumed to be made and received.

    \(\Delta V^{\delta,f,{\cal M}}_t (\omega)\) is now a random variable.
In one common setting , the pertinent parts of the filtration
\({\cal F}_t\) are \({\bf X}_t\) and \({\bf U}_t\). Each possible \(\omega\)
corresponds to (different) realizations \({\bf X}_s\) and \({\bf U}_s\) for \(s
\in [t,t+\delta]\) which in turn lead to realizations of \({\bf CF}\) and
\(V_s\) in that interval.

In market risk (for Market Risk VaR) and counterparty credit risk (for initial
margin requirements), one is interested in quantiles for that random variable,
denoted
\begin{equation}
Q_{\alpha} ( \Delta V^{\delta,f,{\cal M}}_t | {\cal F}_t )
\end{equation}
and computed or approximated as functions
\begin{equation}
q_{\alpha} (t, {\bf X}_t,{\bf U}_t) 
\end{equation}
or
\begin{equation}
q_{\alpha} (t, {\bf R}_t). 
\end{equation}
In the first, one finds the function given the complete
Markovian state. In the second, one finds it as a function 
of a set of regressors ${\bf R}_t$. Often, one tries to 
use as few regressors as possible to minimize computational
requirements, for instance using  only \(V_t\) itself or together
with a few main risk factors.

In the initial margin requirement case, the percentile  \(\alpha\) is typically
1\% (or 99\% if seen from the other side).
For market risk VaR, \(\alpha\) is often 3\%, 1\%, or 0.3\% (or 97\%, 99\%,
99.7\% if seen from the other side).

Often initial margin and market risk value at risk are computed with \(t\)
representing today. However, to understand the impact of future initial margin
requirements or future values at risk, one considers \(t\) representing times
beyond the valuation date to model "future initial margin requirements" (FIM) or
"future VaR". Then one computes future expected margin requirements (or VaR) or
quantiles of the future expected margin requirements (or VaR). The first is
called DIM for future expected margin requirements which can be written as,
\begin{align}
\mathrm{DIM}_{t_0,t,\delta} &= E^\mathbb{P}[Q_\alpha(\Delta
V_t|\mathcal{F}_t)|\mathcal{F}_{t_0}] \nonumber \\
& = E^\mathbb{P}[\mathrm{IM}_{t,\delta}|\mathcal{F}_{t_0}]  \label{timeevolutiondim}
\end{align}

$\mathbb{P}$ would be a pricing measure if the future expected margin requirements 
are computed for pricing purposes or could be a historical measure if computed 
for risk management purposes. 

However, once computed, the initial margin requirements or VaR can be used in
follow up computations. For initial margin requirements, one can compute the
expected cost to fund future initial margin requirements (MVA). VaR (and now
also, Expected Shortfall (ES)) are used in the computation of required
regulatory or economic capital and furthermore the cost to fund it (KVA).
While we do not cover these applications here, the methods described
here can and have been applied to them, and will be described elsewhere.

In this paper, we assume that the stochastic process for \(V_t\) respectively
the functions for \(V\) are given or approximated independently and before
applying the methods given in the paper. In a follow-up paper, we describe how
the \(V_t\) can be computed through backward deep BSDE methods and how to
implement methods to compute the quantiles in that particular case.

The rest of the paper is organized as follows: Section \ref{assumptions} briefly
outlines the assumptions and requirements for the methods studied. Section
\ref{nestedmc} describes the Nested Monte-Carlo procedure to estimate future
initial margin that will be used as a benchmark to compare other methods
against. Section \ref{momentmatching} describes methods using moment-matching
techniques implemented on cross-section of the outer simulation paths, such as
Johnson Least squares Monte-Carlo (JLSMC) and Gaussian Least Squares
Monte-Carlo (GLSMC) in order to fit an appropriate distribution for the
computation of initial margin. Section \ref{quantilereg} describes the quantile
regression method and its implementation. Section \ref{deltagamma} describes the
Delta-Gamma methods which utilize the analytic sensitvities of the instruments
to estimate conditional quantile and future initial margin. Section
\ref{reducedinnersamples} analyzes the effect of adding limited number of inner
samples for estimation of future initial margin and DIM. While the methods
described in the earlier sections with the exception of Nested MC all work on the
cross section of outer paths, the estimation of the properties of the
distribution could be refined further by the addition of limited number of inner
paths, which is the subject of this section. 
In Section \ref{methodcomparison}, we present a comparison of running times and
RMSE across many methods for the call and call combination examples and discuss
their performance. We end with conclusion and discussion in Section
\ref{conclusion}.

\section{Approaches and their Assumptions and Requirements}\label{assumptions}

Estimating future initial margin as well as updating the estimates during
frequent changes in market conditions, changes in counterparty risk profiles and
changes in portfolio itself is an arduous task especially for large and complex
portfolios with many instruments and underliers. With sufficient computational
resources, one can perform simulations of risk factors and cashflows under
several portfolio scenarios. As discussed in the subsequent section, to use
Monte Carlo to estimate the Value at Risk and the quantiles of the portfolio
value change at a future time for each of the several portfolio scenarios, it is
necessary to perform nested simulations at each particular time for each
scenario which further increases the compuational requirements.

Therefore, to avoid nested simulations on large number of portfolio scenarios,
and to estimate future initial margin (FIM) requirements, there are several
approximation methods that compute percentiles and other distribution properties
of the portfolio value change. These methods are given pathwise risk factor values and
portfolio values along simple or nested set of paths either as pathwise values
or as functions of risk factor values. 
 
We briefly review a few of these methods to estimate $\mathrm{FIM}$ under
suitable assumptions, such as whether portfolio values ($V(t)$), portfolio
sensitivities (Delta and Gamma), complete or limited nested paths in the
simulation etc. are available.

We present the procedures, implementations and
comparison of 1) Nested Monte-Carlo, 2) Delta-Gamma and Delta-Gamma Normal, 3)
Gaussian Least Square Monte Carlo (GLSMC), 4) Johnsons Least Square Monte-Carlo
(JLSMC) and 5) Quantile Regression, where each is applicable under certain
assumptions and has certain requirements.

If portfolio values are not given or otherwise computed (pathwise or as a
function) - and thus the methods discussed in this paper cannot be directly
used, contemporary approaches such as DeepBSDE and Differential Machine Learning (DML)
can be extremely useful to compute required portfolio values (and also
first order portfolio sensitivities).
We will discuss Backward DeepBSDE approaches (where portfolio values will be
determined from given sample paths for risk factors and cashflows) and
their use in computing future initial margin requirements in a separate future
paper.
A summary of the various approaches and their requirements is shown
in Table \ref{reqmatrix}.
\begin{table} 
\begin{tabular}{|l|c|c|c|c|c|}\hline
\diaghead{\theadfont Diag ColumnmnHead II}%
{Methods}{Requirements}&
\thead{Risk Factor\\ Simulation\\ and MPoR\\ Cashflows}&\thead{Nested \\Monte-Carlo\\ Paths}&\thead{Portfolio\\Values}&\thead{Sensitivities\\(Delta and\\ Gamma)}&\thead{Limited\\ Nested\\ Paths} \\
\hline 
\makecell{Nested MC} &\checkmark &\checkmark &\checkmark &&\checkmark  \\
\hline
\makecell{Delta Gamma and \\Delta Gamma Normal} &\checkmark &&\checkmark &\checkmark & \\
\hline 
\makecell{JLSMC and \\ GLSMC} &\checkmark &&\checkmark && \\
\hline 
\makecell{Johnson Percentile \\Matching (JPP)} &\checkmark &&\checkmark && \checkmark \\
\hline 
\makecell{Quantile Regression} &\checkmark &&\checkmark && \\
\hline 
\makecell{Backward DeepBSDE\\(Discussed in future paper)} &\checkmark &&&& \\
\hline
\end{tabular}
\caption{Matrix of the available methods to compute VaR and the corresponding
requirements of the methods indicated by \checkmark s. Backward DeepBSDE 
methods also need simulations of all instrument cashflows to learn portfolio 
values and their first order sensitivities.}
\label{reqmatrix}
\end{table}

As noted in the table, different methods are applicable under different
assumptions and have different requirements. This paper aims to
introduce these different methods, provide a comparison of them as well as
highlight their differences and their results obtained for various instruments .

\section{Empirical Nested Monte-Carlo with Many Inner
Samples}\label{nested-monte-carlo}\label{nestedmc}

One particular, "brute force", approach, is to first simulate "outside"
\({\bf X}_t\) (and \({\bf U}_t\)) along a number \(N_O\) of "outside" paths - or
to generate those \({\bf X}_t\) (and \({\bf U}_t\)) distributed according to some "good"
distributions - for a number of times \(t_i\), as in \({\bf X}^i_{t_i}\) (and
\({\bf U}^i_{t_i}\)).
For each outer path $j$ and time \(t_i\), one simulates \(N_I\) inner paths, for
\(N_I\) relatively large, starting now at \({\bf X}^j_{t_i}\) and \({\bf
U}^j_{t_i}\), and computes the corresponding \(\Delta V^{\delta,f,{\cal M}}_t
(\omega)\).
Thus one  has \(N_I\) samples of this random variate and can compute the empirical
quantile, obtaining one quantile for each simulated \({\bf X}^j_{t_i}\) (and
\({\bf U}^j_{t_i}\)), and follow-up computations based on these sampled
quantiles can proceed.

Similarly, one can compute moments from these many inner samples, regress these
empirical moments against appropriate regressors, and perform further follow-up
computations based on these moments (such as moment-matching methods).

In some very special circumstances it might be possible to directly characterize
the distributions and quantiles analytically or approximate them in closed form,
but in the general case, Nested Monte-Carlo is a natural benchmark.
Given the enormous computational requirements (including the requirement to
generate nested paths and generate prices on all of them), this is not a method
that can be applied in production settings in general.

\begin{figure}
    \centering
    \includegraphics[width=.6\linewidth]{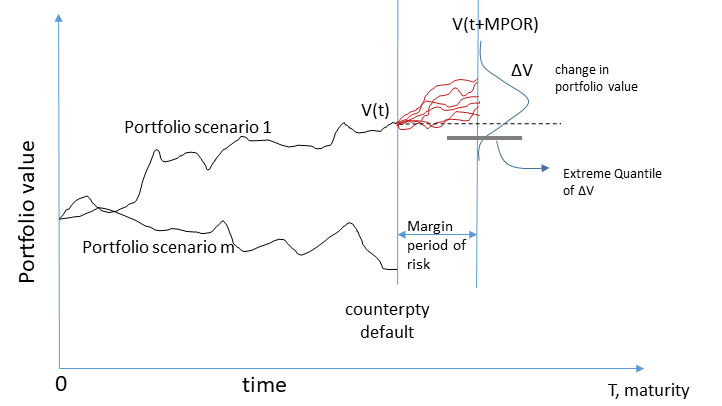}
    \caption{Change in portfolio value as a distribution (or a nested sample) for a single sample path over the Margin Period of Risk.}
    \label{introdimandmpor}
\end{figure}

A Nested Monte-Carlo approach (or, more generally, an approach that looks at
nested distributions) is illustrated in Figure \ref{introdimandmpor}, where the
term ``extreme quantile'' is used to refer to bottom 1\% of the change in
portfolio value.

The method was implemented for a simple Black-Scholes call combination, a
portfolio which consists of a long call at strike $K_1=120.0$ and two short
calls at strike $K_2=150.0$, with a maturity $T=5.0$ years with the
corresponding model parameters  $r=0.03$ and $\sigma=0.1$. The initial spot
price of the underlier was $S_0=85.0$ and the margin period of risk (MPoR) was
fixed at 0.05 years.

\begin{figure}

\begin{subfigure}{.5\textwidth}
  \centering
  \includegraphics[width=.9\linewidth]{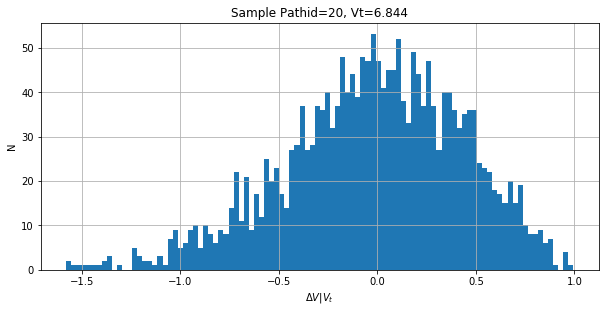}  
  \caption{$\Delta V$ samples for the Call Combination at time t=0.64 years}
  \label{nestedmcsampledist}
\end{subfigure}
\begin{subfigure}{.5\textwidth}
  \centering
  \includegraphics[width=.9\linewidth]{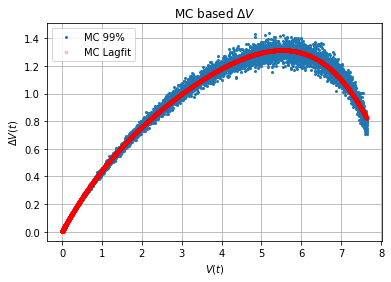}  
  \caption{Conditional 99\% 
  quantiles of $\Delta V$ given $V$ represented by
  blue dots and the corresponding polynomial regressed values shown as red curve.}
  \label{nestedmcconditionalvar}
\end{subfigure}
\caption{Nested Monte-Carlo approach to estimate the Dynamic Initial Margin for the Call Combination.}
\end{figure}

Figure \ref{nestedmcsampledist} shows the samples for $\Delta V_t^{\delta, f,
\mathcal{M}}(\omega)$ for $N_I=1000$ inner samples, for a representative outer
sample path (pathid 20) at time $t=0.64$ years. The portfolio value at that time
for that outer path was $V_t(\omega)=6.844$. Figure \ref{nestedmcconditionalvar}
shows the corresponding 99\% quantiles of $\Delta V$ over the entire cross
section of the outer paths at the same time $t=0.64$ years - the conditional
quantiles $Q_\alpha(\Delta V|V)$.

\section{Nonnested Monte-Carlo: Cross-sectional Moment-Matching Methods}\label{momentmatching}

Due to the high computational requirements for simulation and valuation, it is
rarely possible to spawn nested simulations for inner portfolio scenarios as
required for the Nested MC type estimators, in particular for complex and/or
large portfolios. Often, it is only possible to obtain several independent runs
of the portfolio scenarios. Thus, methods such as Gaussian Least Square
Monte-Carlo (GLSMC) and Johnson Least Square Monte-Carlo (JLSMC)
\cite{mcwalter2018, caspers2017} are used to extract information about the
distribution of the portfolio value change only from such outer simulations.

\begin{figure}
    \centering
    \includegraphics[width=.6\linewidth]{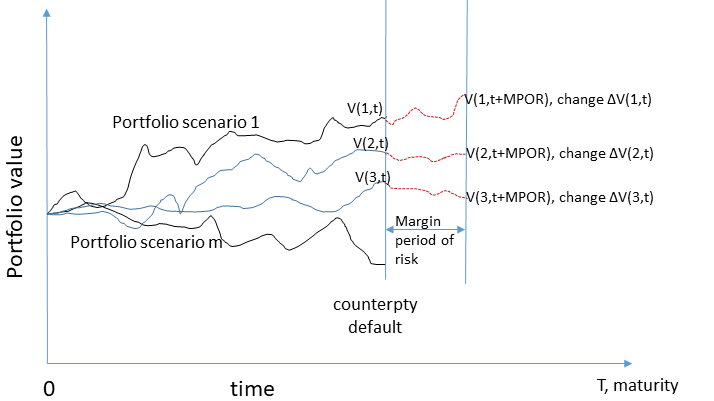}
    \caption{Change in portfolio values from outer simulations is used to
    estimate properties of the distribution $\Delta V$.}
    \label{cleanchangepv}
\end{figure}

The data usually available from outer simulations is illustrated in Figure \ref{cleanchangepv}.

\subsection{Moment-Matching Methods: GLSMC and JLSMC}

Moment-Matching Least Square Monte Carlo methods can be outlined as follows:
\begin{itemize}
\item Generate $M$ portfolio scenarios (outer simulations) corresponding to the
$k$ underlying assets, given the various risk factors and correlations from
time $0$ till maturity  $T$.
\item Define the appropriate change in portfolio value corresponding to sample
path $m$ as,
\begin{equation}
\Delta V_m(t, \delta) = V_m(t+\delta) +\mathrm{DCF}_m(t, t+\delta) - V_m(t) \label{deltavpathm}
\end{equation}
where $V_m$ is the pathwise portfolio value in scenario $m$, and
$\mathrm{DCF}_m$ is the value of the fully or partially included net cashflow
between the counterparties from time $t$ to $t+\delta$.
\item Based on the distribution(s) to be fitted, compute the first four
 raw sample moments  $\Delta V_m,$ $(\Delta V_m)^2$, ($\Delta V_m)^3$,
$(\Delta V_m)^4$ (for Johnson) or the first two raw sample moments
$\Delta V_m$, $(\Delta V_m)^2$ (for Gaussian) that will be used to estimate the
actual raw moments.
\item Accordingly estimate the first four raw moments, $r_1, r_2,
r_3, r_4$, or first two raw moments  $r_1, r_2,$ as a function of the underlying
states $\mathbf{X}_t$ or the portfolio value $V_t$ (or other regressors 
$\mathbf{R}_t$), by a suitable parameteric function such as polynomial 
regression,
\begin{eqnarray*}
r_i &=& f(\bm{\beta}, \mathbf{X}) \approx E[(\Delta V)^i|\mathbf{X}_t = \mathbf{X}] \\
&&\mbox{ or } \\
r_i &=& f(\bm{\beta}, V) \approx E[(\Delta V)^i|V_t = V] \\
\end{eqnarray*}
where $\bm{\beta}$ is a vector of polynomial regression coefficients and $i=1,2,3,4$.

\item GLSMC \cite{caspers2017}: The estimated raw moments $r_1, r_2$ can be used
to approximate the mean and variance of the distribution $\Delta V_t$ (
portfolio value change of interest during MPoR) for an assumed normal distribution of
$\Delta V_t$. The normal distribution with the estimated mean and variance is
then used to estimate the conditional quantile.

\item JLSMC \cite{mcwalter2018}: The estimated first four raw moments $r_1, r_2,
r_3, r_4$ can be used to fit an appropriate Johnson's distribution via Moment Matching
method \cite{hill1976} with (approximately) these first four
moments. The so obtained Johnson's distribution is then used to estimate the 
conditional quantile.
\end{itemize}

\begin{figure}
    \centering
    \includegraphics[width=.6\textwidth]{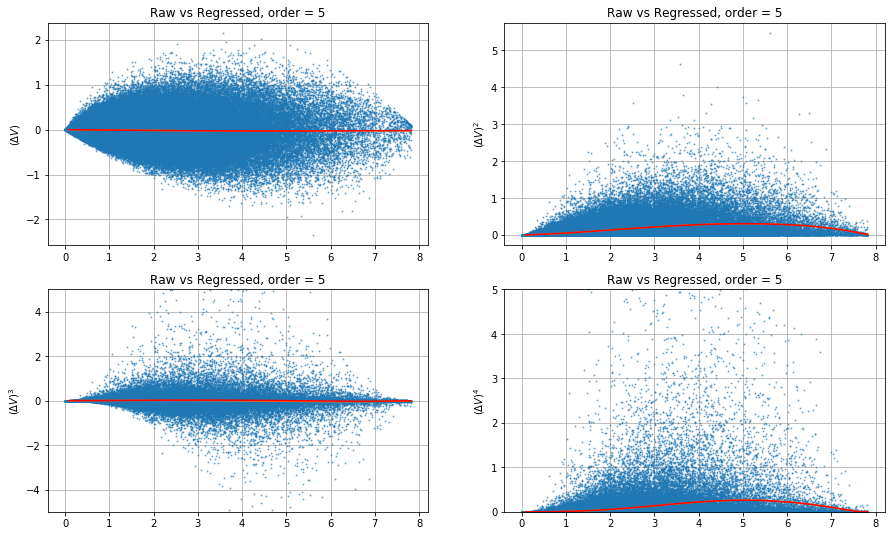}
    \caption{Polynomial regression of the first four raw-moments of the 
    portfolio value change. Plotted over $V_t$.}
    \label{rawvsregressed}
\end{figure}

\begin{figure}
    \centering
    \includegraphics[width=.6\textwidth]{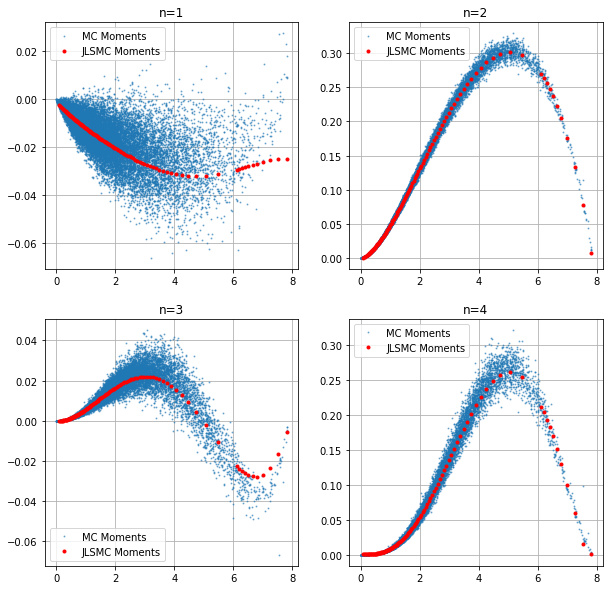}
    \caption{Comparison of the central moments computed from Nested Monte-Carlo
    simulation (blue dots) with those computed from regressed raw moments (red dots). 
    Plotted over $V_t$}
    \label{centralmoments}
\end{figure}

An example of raw moments from the portfolio value change and the corresponding
regressed moments with portfolio value $V_t$ via polynomial regression for the
first four moments for Call Combination instrument is shown in Figure
\ref{rawvsregressed}. Furthermore, the central moments obtained from these raw
moments above (used for both GLSMC and JLSMC) were compared with those obtained
from Nested Monte-Carlo method with $N_I=1000$ inner simulation paths for the
same instrument. The comparison is shown in Figure \ref{centralmoments} and the
central moments obtained via regressed raw moments are seen to be in good
agreement with the Nested MC.

A detailed procedure and the settings used for JLSMC is presented in
\cite{mcwalter2018} and is not repeated here for the sake of brevity. However,
it is noted here that the moment matching algorithm \cite{hill1976} originally
designed for smaller sets of observations is quite involved. Therefore the total
number of given sets of moments that must be fit via the Johnson moment matching
algorithm is equal to the number of outer simulation paths, $N_O=100,000$ at
each time-step where Value at Risk is required. This poses a heavy computational
burden in the absence of additional redesign and parallelization of the moment
matching algorithm. Hence in order to reduce the computational load and obtain
results with comparable accuracy, only moments at select values of the outer
simulation values, either certain values of the underlying risk factors
$\mathbf{X}_t$ or certain quantiles of the portfolio values $V_t$ at time $t$
are used for moment matching. Let $v_j = Q_{j/N}(V_t), j=0,\cdots, N$ be the
$j/N$th quantile of the samples $V_t$ at time $t$ and
\begin{align*}
r^i_j &= f_i(\bm{\beta}, v_j), i=1, 2, 3, 4 \\
&\mbox{ or } \\
r^i_j &= f_i(\bm{\beta}, x_j), i=1, 2, 3, 4 \\
\end{align*}
be the corresponding raw moments obtained via parameteric regression at the
specific values of $V_t$ or $\mathbf{X}_t$. The parameters of the corresponding
Johnson's distribution obtained by the moment matching procedure are given by,  
\begin{equation*}
\mathtt{xi(j), lambda(j), delta(j), gamma(j), type(j)} = \mathtt{MomentMatching(r^1(j), r^2(j), r^3(j), r^4(j))}
\end{equation*}
where $\mathtt{xi(j), lambda(j), delta(j), gamma(j)}=\xi_j, \lambda_j, \delta_j,
\gamma_j$  are the parameters of the  Johnsons distribution, and
$\mathtt{type(j)}$ represents the type or family of the distribution. The pdf of
Johnson's distributions belonging to different families is given by
\cite{johnson1949, mcwalter2018}:
\begin{itemize}
\item $S_L$ family,
\begin{equation}
p(y) = \frac{\delta}{\sqrt{2\pi}}\frac{1}{y}\exp\left\{-\frac{1}{2}[\gamma+\delta \log(y)]^2\right\}, \xi< X<\infty
\end{equation}
\item $S_B$ family,
\begin{equation}
p(y) = \frac{\delta}{\sqrt{2\pi}}\frac{1-y}{y}\exp\left\{-\frac{1}{2}[\gamma+\delta \log(y/(1-y))]^2\right\}, \xi< X<\xi+\lambda
\end{equation}
\item $S_U$ family,
\begin{equation}
p(y) = \frac{\delta}{\sqrt{2\pi}}\frac{1}{\sqrt{y^2+1}}\exp\left\{-\frac{1}{2}[\gamma+\delta \log(y+\sqrt{Y^2+1})]^2\right\}, -\infty< X< \infty
\end{equation}
\end{itemize}
for $y=\frac{X-\xi}{\lambda}$.


\begin{figure}
    \centering
    \includegraphics[width=\textwidth]{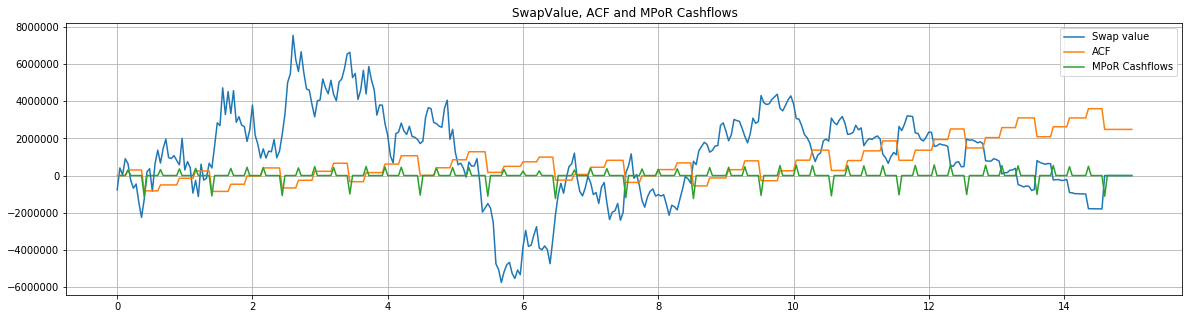}
    \caption{Sample scenario of Interest Rate Swap with the portfolio
    value, accumulated fixed and floating cashflows and cashflows
    within MPoR. Plotted over time.}
    \label{irswapcashflows}
\end{figure}

\textbf{Interest Rate(IR) swap example:} The above methods were compared for a
realistic IR swap example, with one party making annual fixed rate payments with
the counterparty making quarterly floating rate payments over a 15 year period.
Here Party A pays floating rate with a constant spread of 0.9 \% every 3 months
over the current floating interest rate (LIBOR/PRIME etc). Party B pays fixed
annual rate of 4.5 \% every year based on the interest rate agreed at the start
of the contract. The outstanding notational varies from 36.2 million dollars in
the first year to 63.2 million dollars at the end of year 15. The Margin Period
of Risk is set to 10 business days or 0.04 years, with a total of 375 intervals
until maturity. The interest rate and risk factors were simulated following G1++
process and a sample path for cashflows, portfolio values and the risk factors
is shown for a sample scenario in Figure \ref{irswapcashflows}.

\begin{figure}
    \centering
    \includegraphics[width=0.5\textwidth]{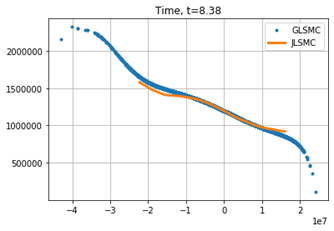}
    \caption{Comparison of Conditional quantiles at $\alpha=0.01$ obtained via GLSMC
    and JLSMC. The range of the JLSMC curve is smaller as only the quantiles 1
    through 99 of values of $V_t$ at time $t=8.38$ were considered for moment
    matching and for $\Delta V$ quantile estimation. Plotted over $V_t$.}
    \label{irswapcondvarglsvsjls}
\end{figure}
The conditional 1\% quantiles obtained via GLSMC and JLSMC at time $t=8.38$
years are compared in Figure \ref{irswapcondvarglsvsjls} and are found to be in
good agreement. The GLSMC method is implemented over the entire cross-section of
the outer paths, whereas the JLSMC method, as discussed earlier, was implemented
at specific quantiles 1 through 99 of the values of $V_t$. 

\begin{figure}
\begin{subfigure}{.5\textwidth}
  \centering
  \includegraphics[width=.9\linewidth]{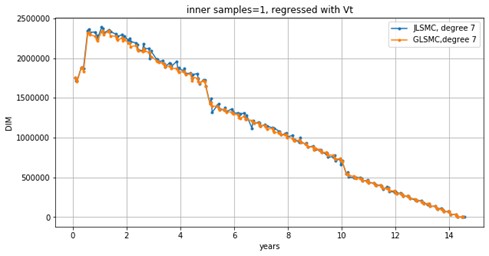}  
  \caption{Comparison of time evolution of DIM obtained via GLSMC and JLSMC
  with MPoR cashflows.}
\end{subfigure}
\begin{subfigure}{.5\textwidth}
  \centering
  \includegraphics[width=.9\linewidth]{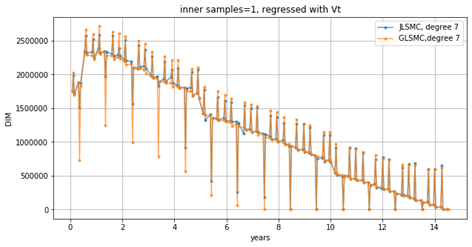}  
  \caption{Comparison of time evolution of DIM obtained via GLSMC and JLSMC
  without MPoR cashflows.}
\end{subfigure}
\caption{Time Evolution of DIMs for the IR Swap with moments
regressed via Laguerre polynomials of degree 7.}\label{irswaptimedim}
\end{figure}

Finally the time evolution of DIM for the same instrument over the entire 15
year period as defined in Equation \eqref{timeevolutiondim} was compared for
GLSMC and JLSMC. The results are shown in Figure \ref{irswaptimedim} where the
first four moments were regressed with Laguerre polynomials of degree seven.
On the left, cashflows during MPoR are included in the change of value, on the
right, MPoR cashflows are not included. It can be seen that not assuming that
all cashflows are paid can make the margin requirements much more spiky. The
results for the different methods are seen to be in good agreement with each
other.

\subsection{Further Analysis: JLSMC Moment Regression}

We present another instrument, namely a cross currency FX-Call option in order
to highlight the differences in the results obtained via the GLSMC and JLSMC
methods. The Black-Scholes FX Call is simulated and priced with a constant
domestic risk-free rate $r_d=0.08$, and a constant foreign rate of $r_f=0.02$
along with FX rate volatility of $\sigma=0.3$. The current spot FX rate is
$S_0=100.0$.  The option has maturity $T=1.0$ year and strike $K=105.0$. Similar
to the other examples, the margin period of risk is fixed at $0.04$ years, with
25 MPoR intervals till maturity.

An algorithm to fit an appropriate Johnson's distribution (Sl - logarithmic,
Sb - bounded, Su -unbounded and Sn - normal) given the first four raw or central
moments is presented in \cite{hill1976}. Raw moments are estimated by
 independent polynomial regression on the first four powers of the
change in portfolio values as shown in Figure \ref{rawvsregressed}.

Johnson's analysis of the types of distributions \cite{johnson1949} also
outlines the constraints for various moments, under which a valid distribution
can be found. The relationship between Skew ($\beta_1$) and Kurtosis ($\beta_2$)
requires that $\beta_2\geq \beta_1 +1$. However, since the moments are regressed
independently without imposing any joint constraints, there is no guarantee that
relationship between Skew and Kurtosis will be satisfied everywhere when the
regressed form of the moments is used.
In fact, it can be seen that in some examples under some regression models, the
Skew-Kurtosis relationship fails to hold at some points, and therefore one is
unable to find a distribution that matches the moments, as shown in Figure
\ref{polynomialmomentregressioncorrection}. This is elaborated further in the
following subsection.

\subsubsection{Polynomial Moment Regression}

\begin{figure}
\begin{subfigure}{\textwidth}
    \centering
    \includegraphics[width=0.5\textwidth]{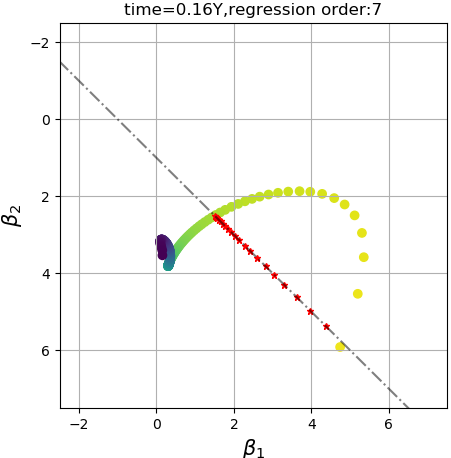}
    \label{momentregressioncorrection}
    \caption{The skew-kurtosis relationship.}
\end{subfigure}
\begin{subfigure}{\textwidth}
    \centering
    \includegraphics[width=\textwidth]{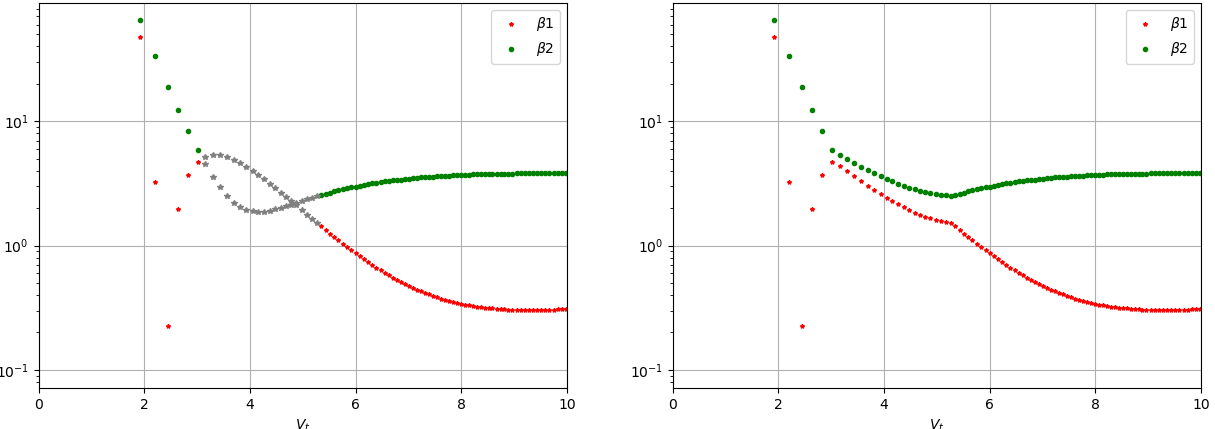}
    \label{b1b2correction_vs_vt}
\end{subfigure}
\caption{{\small Skew-Kurtosis obtained from independently regressed moments is
shown at selected quantiles (1 through 99) of $V_t$ (outer simulation) at time
$t=0.16$ years.  (Top) The datapoints that lie above $\beta_2=\beta_1+1$ (dotted
line) violate the skew-kurtosis relationship and hence do not correspond to a
valid distribution. (Bottom) Corresponding raw $\beta_1$ and $\beta_2$ against
$V_t$ for the same time. (Bottom-Left) The gray datapoints denote those
violating the relationship, (Bottom-Right) The moment-corrected data     
points.}}\label{polynomialmomentregressioncorrection}
\end{figure}

The Skew vs Kurtosis obtained from the regressed raw moments at a single time is
 shown in Figure \ref{polynomialmomentregressioncorrection}. Each data point 
corresponds to a unique path with a portfolio value $V_m(t)$ and its
corresponding portfolio value change $\Delta V_m$. For simplicity only equally
spaced quantiles from 0.01 to 0.99 of $V(t)$ are shown. The darker colors
indicate lower quantiles closer to 0.01 and lighter colors indicate higher
quantiles closer to 0.99, along with the line $\beta_2 = \beta_1 +1$. The data
points where the condition is satisfied appear below that line and those that do
not satisfy the constraint, appear above the $\beta_2 = \beta_1 +1$ line (the
$\beta_2$ axis is reversed).

In the standard implementation of the moment-matching algorithm
\cite{hill1976,jones2014}, the data points not satisfying the relationship are
discarded or just replaced by normal distribution with the given mean and
variance only (first and second moments only) thus essentially reverting to the
procedure for GLSMC in that case.

\begin{figure}
\begin{subfigure}{.5\textwidth}
  \centering
  \includegraphics[width=\linewidth]{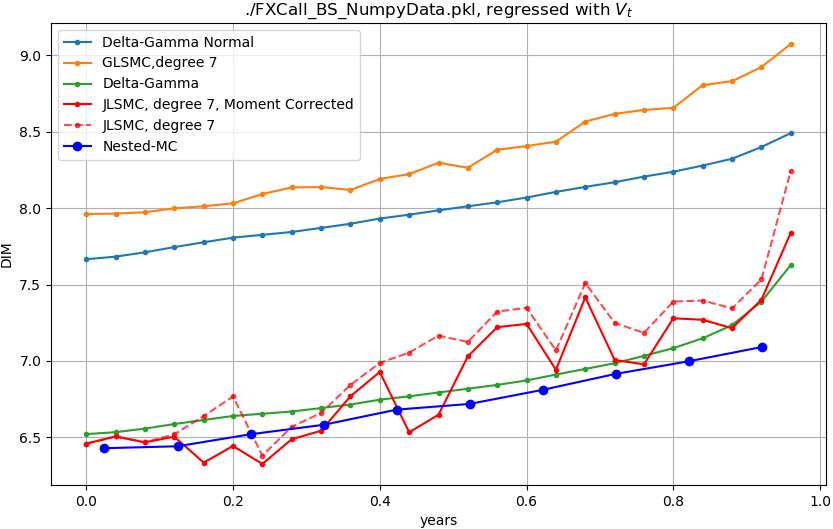} 
  \caption{Comparison of time evolution of DIM obtained via GLSMC and JLSMC with
  polynomial regression of order 7 (with and without moment correction) for
  the FX option.}
  \label{subfig:timedimfxoptions}
\end{subfigure}
\begin{subfigure}{.5\textwidth}
  \centering
  \includegraphics[width=\linewidth]{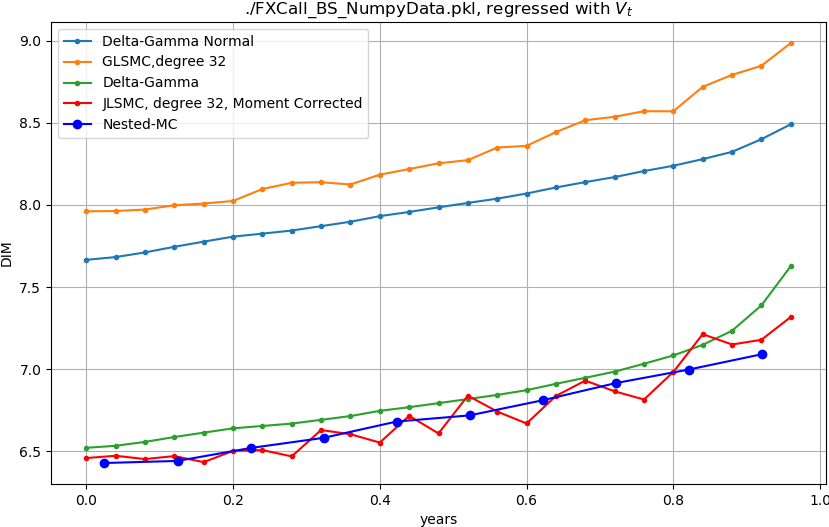}  
  \caption{Comparison of time evolution of DIM obtained via GLSMC and JLSMC for
  the FX option with moment correction and polynomial order 32.}
  \label{subfig:timedimfxoptionswcorrects}
\end{subfigure}
\caption{Time Evolution of DIM for FX options with and without moment
correction.}\label{timedimmomentcorrection}
\end{figure}

The results of the time evolution of DIM for the FX option is shown in Figure
\ref{timedimmomentcorrection}. It can be seen that although various methods
agree well for the IR swap example (Figure \ref{irswaptimedim}), they deviate
from each other for the FX option.
\begin{itemize}
\item This strongly indicates that different instruments have their own 
characteristic distributions for $\Delta V(t)|\mathcal{F}_t$ and hence cannot
be approximated by normal distribution alone with the given mean and variance
as in the GLSMC method. This also shows that a more accurate
characterization of the distributions is needed, as in the JLSMC method.
\item As mentioned earlier, when the Skew-Kurtosis relationship is not satisfied
for the regressed moments, either the point is discarded or only a normal
distribution matching only two moments is used. For the FX option, with   the
given polynomial regression, more data points violate the Skew-Kurtosis
relationship at later times $t$. As those points are modeled by normal
distributions rather than more expressive distributions, the estimated trends
for DIM by JLSMC converge to    that of GLSMC for such later times $t$ as as
shown in Figure \ref{subfig:timedimfxoptions}, which is clearly an artifact of 
the standard implementation.
\end{itemize}

\textbf{Polynomial Regression Correction:} To avoid that data points are
discarded or only represented by normal distributions, it is possible to map the
invalid Skew and Kurtosis value pairs to the closest Skew and Kurtosis value
pair that does not violate the relationship and use the corresponding Johnson
distribution. This is illustrated in Figure
\ref{polynomialmomentregressioncorrection} where the data points above the
$\beta_2 = \beta_1 + 1$ line are projected onto the line in a (orthogonal)
least-square sense as indicated by the red points. This ad-hoc method provides a
better approximation of the distribution via a valid Johnson's distribution
instead via a normal distribution. In addition, a better model with a higher
order polynomial regression or better regresion techniques such as ensemble
learning via Random Forest might mitigate the issue with fewer data points
violating the moment constraints. The time evolution of DIM with the proposed
moment corrections is shown in Figure \ref{subfig:timedimfxoptionswcorrects}.

\section{Nonnested Monte-Carlo: (Conditional) Quantile
Regression}\label{quantilereg}

Conditional quantile regression \cite{koenker2001} can be used to estimate
population quantiles based on observations of a dataset $(\mathbf{C}_i, y_i)$
where $\mathbf{C}_i$ for each $i$ is a vector of covariates or features and 
$y_i$ is the quantity of interest (for instance, low birth weight or CEO
compensation depending on maternal or firm characteristics, respectively,
in two examples in \cite{koenker2001}). Based on the vector $\mathbf{R}_i$ of
covariates one regresses against, one estimates $q_{\alpha}(Y|\mathbf{R})$. The
quantile regression procedure along with elicitability criteria for loss
functions was used to extract conditional VaR and Expected Shortfall (ES) for
XVA applications in \cite{albanese2020xva}.   

We will use conditional quantile regression \cite{koenker2001} as follows:
Assuming that we have given ($r$-dimensional) samples $\mathbf{R}_i$ of
regressors (chosen from a larger set of possible features) and samples $Y_i$
from the (one-dimensional) conditional distribution to be analyzed, the function
$f_{\alpha}(\mathbf{R})$ is characterized as the function from an appropriately
parametrized set of functions that minimizes the loss function
\begin{equation}
L_\alpha = \sum_i \rho_{\alpha} (Y_i - f_{\alpha}(\mathbf{R}_i)) 
\end{equation}
with 
\begin{equation}
\rho_{\alpha} (x) =\left\{ 
\begin{array}{ccc}
\alpha x  & \mbox{ if } x \geq 0 \\
(\alpha-1) x & \mbox{ if } x < 0
\end{array}
\right. .
\end{equation} 

$f_\alpha$ defined by $\argmin L_\alpha$ then estimates the conditional quantile
$q_{\alpha}(Y|\mathbf{R})$.
\cite{koenker2001} consider parametric functions $\mu(\mathbf{R},\beta)$ and in
particular, functions that depend linearly on the parameters $\beta$.
In their case, the optimal parameters can then be computed by linear
programming.
Here, we will apply conditional quantile regression both in the linear and
nonlinear approximation context.

In our application, the conditional target distribution is the one of the
portfolio value change  $Y_m=\Delta V_m$ from time $t$ to time $t+\delta$ given
some regressors or the complete Markovian state. The covariates that can be used
have to be measurable as of time $t$. Assuming that the portfolio value can be
exactly or approximately be considered Markovian in terms of current risk factor
values $\mathbf{X}_t$ and extra state $\mathbf{U}_t$, $\mathbf{X}_t$ and
$\mathbf{U}_t$ should contain all necessary information. Computing conditional
quantiles conditional on the full state might be too expensive and thus, a
smaller set of covariates might be used, such as the portfolio value $V_t$ and
some ``most important risk factors''.

In our machine and deep learning setting, we thus numerically minimize the loss 
function given above within an appropriate set of functions depending on the 
chosen regressors.

It is important to be parsimonious in the selection of the model and the number
of free parameters, since with sufficient number of free parameters (and if
there is only one value of $\Delta V_m$ for each $\mathbf{R}_m$), it is possible
to learn each $\Delta V_m$  for each $\mathbf{R}_m$.  A simple polynomial
regression or a sufficiently simple neural network can strike a balance between
a reasonably smooth approximation of the quantiles and achieving a local minimum
for the loss function.

\begin{figure}
\begin{subfigure}{0.9\textwidth}
  \centering
  \includegraphics[width=\linewidth]{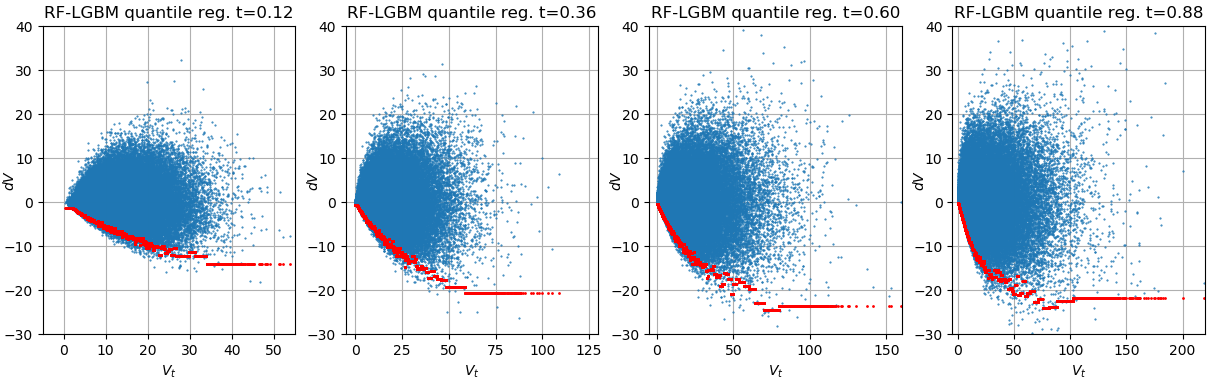}  
  \caption{Quantile regression via RandomForest-LightGBM on clean portfolio
  changes at various times.}
\end{subfigure}\\
\begin{subfigure}{0.9\textwidth}
  \centering
  \includegraphics[width=\linewidth]{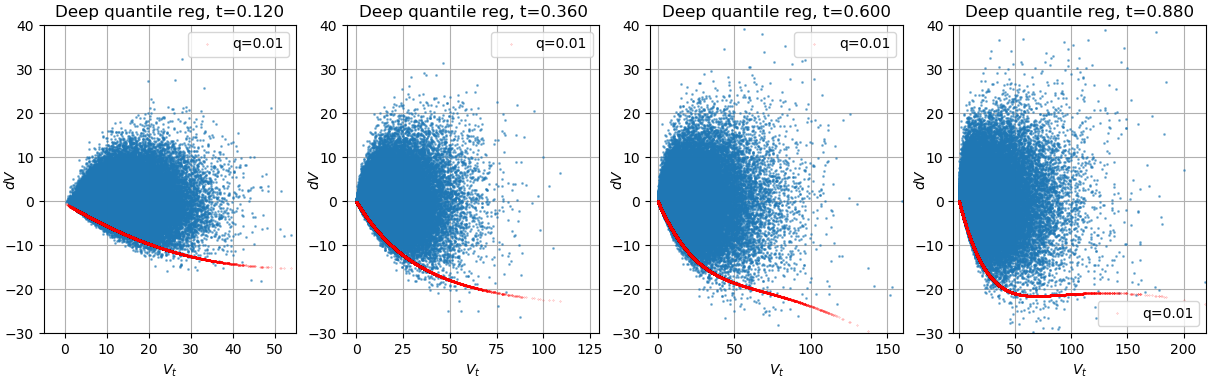}  
  \caption{Quantile regression via Deep Neural Networks on clean portfolio
  changes at various times.}
\end{subfigure}
\caption{Quantile regression performed on the clean change in portfolio values
for the given outer portfolio scenarios for FX option
instrument.}\label{quantileregcondvar}
\end{figure}

Figure \ref{quantileregcondvar} shows results of quantile regression for $dV|V$
with RF-LightGBM (upper panel) and DNN (lower panel). In the body of the $V_t$
distribution, results seems to mostly agree (with DNN giving a smoother results
rather than the step wise curve coming from RF-LightGBM). For larger $V_t$, the
behavior seems to be somewhat but not completely different.

Similarly as in least square regression, assuming that one tries to determine a
quantile function as a linear combination of given functions of given features,
there are faster deterministic numerical linear algebra methods available rather
than resorting to generic optimization methods - linear least squares by SVD (in
least square regression) and linear programming (for quantile regression). Once
one no longer operates in the space of linear combinations of given functions,
one needs to apply more generic and general numerical optimization techniques
such as (stochastic) gradient descent or similar variants, both for least square
regression and quantile regression.

\begin{figure}
\begin{subfigure}{0.9\textwidth}
  \centering
  \includegraphics[width=\linewidth]{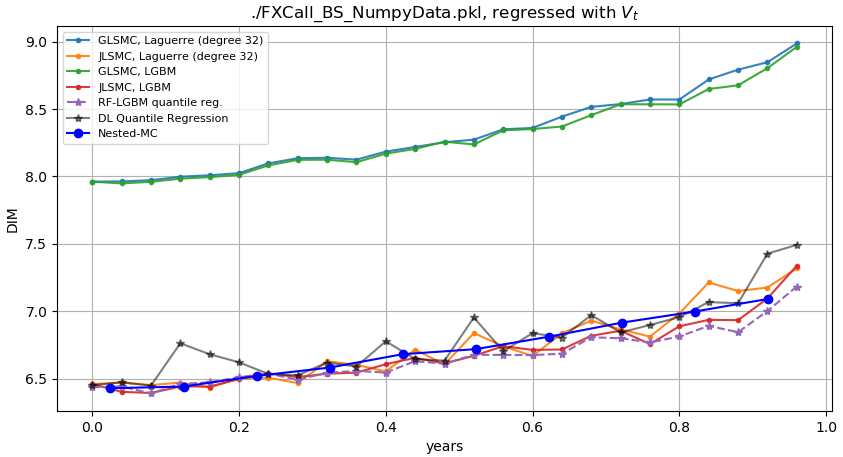}  
  \caption{Comparison of various methods for the time evolution of DIM for FX
  option instrument.}
\end{subfigure}\\
\begin{subfigure}{0.9\textwidth}
  \centering
  \includegraphics[width=\linewidth]{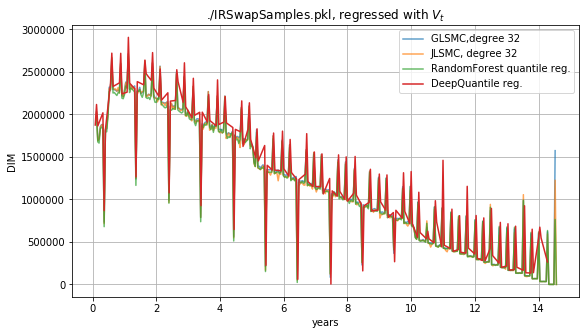}  
  \caption{Comparison of various methods of the time evolution of DIM for IR
  swap instrument.}
\end{subfigure}
\caption{Comparison of time-evolution of DIM for the FX option and IR swap
instruments.}\label{quantileregdim}
\end{figure}

Figure \ref{quantileregdim} compares GLSMC, JLSMC, and Quantile Regression
methods for the FX option (upper panel) and IR swap (lower panel). For the FX
option, JLSMC and Quantile Regression seem to mostly agree, differing somewhat
in smoothness (and looking similar to the earlier Nested MC results) while GLSMC
gives materially different (and larger) values. For IR swap, all give broadly
similar results.

\section{Nonnested Monte-Carlo but Given Sensitivities}\label{deltagamma}
Another technique to estimate the change in portfolio values over MPoR is to use
portfolio sensitivities if available. The distribution of the return of the
underlying risk factors can be combined with the portfolio sensitivies,
specifically the first and second derivatives (portfolio Delta and Gamma) to
obtain a good approximation of the distribution of the change in portfolio
values over MPoR.

Let $\mathbf{S}(t) = (S_1(t), \cdots, S_k(t))$ be the values of the $k$ underliers
that constitute the given portfolio and let $\mathbf{R}(t) = (\Delta S_1/S_1,
\cdots, \Delta S_k/S_k)$ be the corresponding vector of asset returns where 
$\Delta S_i(t) = (S_i(t+\Delta t)- S(t))$.
The second-order approximation of the change in the value of portfolio can be
written in terms of the sensitivities as \cite{book:shreve04,book:alexander08},
\begin{equation}
\Delta V \approx \bm{\delta}_\$ \mathbf{R} + \frac{1}{2} \mathbf{R}^T\mathbf{\Gamma}_\$\mathbf{R}
\end{equation}
where,
\begin{eqnarray}
\bm{\delta}_\$ &= &(\delta_1^\$, \delta_2^\$, \cdots, \delta_k^\$)^T \\
\delta_i^\$ &= &\frac{\partial V}{\partial S_i}\times S_i \\
\mathbf{\Gamma}_\$ &= &(\gamma_{ij}^\$) \\
\gamma_{ij}^\$ &= &\frac{\partial^2 V}{\partial S_i\partial S_j}\times S_i \times S_j
\end{eqnarray}
where $V$ is the value of the portfolio corresponding to the value of the
underliers. For a single underlier, the above equation can be written as,
\begin{equation}
\Delta V \approx \delta^\$ R + \frac{1}{2} \gamma^\$ R^2 \label{singledeltav}
\end{equation}
for corresponding single asset return $R$. 

\subsection{Delta-Gamma Normal}
As a first approximation, the change in portfolio values can be considered to be
normally distributed, with mean and variance obtained from Eq.
(\ref{singledeltav}). Here the discounted asset return $R$, is assumed to be
normally distributed with mean zero and variance $\Omega=\sigma^2$, (i.e) $R
\sim \mathscr{N}(0,\Omega)$.
\begin{align*}
\Omega_h &= \sigma^2 \Delta t \\
\mathrm{E}(\Delta V) &= \frac{1}{2} \mbox{tr}(\Gamma_\$ \Omega_h) \\
\mbox{Var}(\Delta V) &= \frac{1}{2} \mbox{tr}[(\Gamma_\$ \Omega_h)^2] + \delta'_\$ \Omega_h \delta_\$
\end{align*}
further with the approximation that $\Delta V$ follows a normal distribution,
the corresponding $\alpha$th quantile of the distribution can be written as, 
\begin{equation}
Q_\alpha(\Delta V|V) = z_\alpha \sqrt{(\mbox{Var}(\Delta V))} + \mathrm{E}(\Delta V)
\end{equation}
where $z_\alpha$ is the $\alpha$th quantile of the standard normal distribution.

\begin{figure}
    \centering
    \includegraphics[width=.6\linewidth]{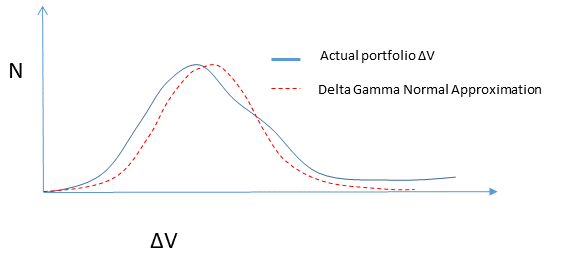}
    \caption{Approximation of $\Delta V$ distribution by Delta-Gamma
    Normal approach could lead to deviation from the actual $\Delta V$
    distribution.}
    \label{normalvsactualdeltaV}
\end{figure}

However this approximation, which could significantly deviate from the actual
$\Delta V|V$ distribution, could lead to over-estimation or under-estimation of
the margin requirements. This deviation due to the approximation is apparent
from the actual sample distribution shown in Figure \ref{nestedmcsampledist},
which is skewed and could be heavy tailed dependending on value of the portfolio
and the risk factors (Figure \ref{normalvsactualdeltaV}). Comparison of DIM
obtained via this approximation against other methods for the call combination
discussed earlier is shown in Figure \ref{comparevardeltagamma} and Figure
\ref{comparedimdeltagamma}. Hence although the normal approximation of the
$\Delta V|V$ distribution allows for rapid estimation of initial margin
requirements, it may be not be suitable for complex portfolios or portfolios
with more complex instruments.

\subsection{Delta-Gamma with Cornish-Fisher Cumulant Expansion}
A further refinement of the previous method is to relate the quantiles of an
empirical distribution to the quantiles of standard normal distribution via the
higher order moments of the empirical distribution and the so called
Cornish-Fisher expansion. Given the raw moments, and the corresponding central
moments, ($cm_1, cm_2, cm_3, cm_4, cm_5$) from raw moments, compute the
skew ($\kappa_3$), kurtosis ($\kappa_4$) and $\kappa_5=cm_5^5/cm_2^{5/2}$ from
the central moments. Hence, the Delta-Gamma method approximates the quantiles of
the portfolio value changes $\Delta V|V$ in terms of its raw moments and uses the
Cornish-Fisher expansion to compute the $\alpha$th percentile of $\Delta V|V$.
\begin{align}
Q_\alpha(\Delta V|\mathcal{F}_t) &= z_\alpha + \kappa_3\frac{(z_\alpha^2 - 1)}{6} + \kappa_4\frac{z_\alpha^3 - 3z_\alpha}{24} \nonumber \\
&- \kappa_3^2 \frac{(2z_\alpha^3-5z_\alpha)}{36} + \kappa_5\frac{z_\alpha^4 - 6z_\alpha^2 + 3}{120} \nonumber \\
&- \kappa_3\kappa_4\frac{z_\alpha^4-5z_\alpha^2+2}{24} + \kappa_3^3\frac{12z_\alpha^4-53z_\alpha^2 + 17}{324} \label{cfexpansion}
\end{align}
where $z_\alpha$  is the $\alpha$th quantile of a standard normal distribution and
$\kappa_3, \kappa_4, \kappa_5$ are the skew, kurtosis and higher order moments
of the distribution. The Delta-Gamma approach approximates the above quantities
from the corresponding portfolio sensitivities and asset returns over the MPoR.

Similar to the Delta Gamma Normal method, the discounted asset return is assumed
to be normally distributed with mean zero and variance $\Omega$, (i.e) $R \sim
\mathscr{N}(0,\Omega)$. The corresponding raw moments written as powers of
$\Delta V$ in terms of $\delta^\$(=\delta), \gamma^\$(=\gamma), R$, and $\Omega$
are,
\begin{eqnarray}
\Delta V &=& \delta^\$ R + \frac{1}{2} \gamma R^2 \label{deltav1}\\
\Delta V^2 &=& \delta^2 R^2 + \frac{1}{4}\gamma^2 R^4 + \delta \gamma R^3 \label{deltav2}\\
\Delta V^3 &=& \delta^3 R^3 + \frac{1}{4}\delta \gamma^2 R^5 + \delta^2 \gamma R^4 + \frac{1}{8} \gamma^3 R^6 + \frac{1}{2}\delta \gamma^2 R^5 + \frac{1}{2}\delta^2 \gamma R^4 \nonumber \\ 
&=& \delta^3 R^3 + \frac{3}{4}\delta \gamma^2 R^5 + \frac{3}{2}\delta^2\gamma R^4 + \frac{1}{8}\gamma^3R^6 \label{deltav3}\\
\Delta V^4 &=& \delta^4 R^4 + \frac{3}{4}\delta^2 \gamma^2 R^6 + \frac{3}{2}\delta^3 \gamma R^5 + \frac{1}{8}\delta \gamma^3 R^7 + \frac{3}{4}\delta^2 \gamma^2 R^6 + \frac{1}{2}\delta^3 \gamma R^5 + \frac{3}{8}\delta \gamma^3 R^7 + \frac{1}{16}\gamma^4 R^8 \nonumber \\
&=& \delta^4 R^4 + \frac{3}{2} \delta^2 \gamma^2 R^6 + 2\delta^3\gamma R^5 + \frac{1}{2}\delta \gamma^3 R^7 + \frac{1}{16}\gamma^4 R^8 \label{deltav4}\\
\Delta V^5 &=& \delta^5 R^5 + \frac{3}{2}\delta^3 \gamma^2 R^7 + 2\delta^4 \gamma R^6 + \frac{1}{2}\delta^2 \gamma^3 R^8 + \frac{1}{16}\delta \gamma^4 R^9 \nonumber \\
&& + \delta^3 \gamma^2 R^7 + \frac{1}{2}\delta^4 \gamma R^6 + \frac{3}{4}\delta^2 \gamma^3 R^8 + \frac{1}{4}\delta \gamma^4 R^9 + \frac{1}{32} \gamma^5 R^{10} \nonumber \\
&=& \delta^5 R^5 + \frac{5}{2} \delta^3 \gamma^2 R^7 + \frac{5}{2} \delta^4 \gamma R^6 + \frac{5}{4}\delta^2 \gamma^3 R^8 + \frac{5}{16} \delta \gamma^4 R^9 + \frac{1}{32} \gamma^5 R^{10} \label{deltav5}
\end{eqnarray}
Using the first 10 moments of the normal distribution, the  expectations of the
above powers of $\Delta V$ can be written as 
\begin{align*}
\mathrm{E}(\Delta V) &= \frac{1}{2}\gamma \Omega \\
\mathrm{E}(\Delta V^2) &= \delta^2 \Omega + \frac{3}{4} \gamma^2 \Omega^2 \\
\mathrm{E}(\Delta V^3) &= \frac{3}{2}\delta^2 \gamma (3\Omega^2) + \frac{1}{8}\gamma^3(15\Omega^3) \\
\mathrm{E}(\Delta V^4) &= \delta^4 (3\Omega^2) + \frac{3}{2} \delta^2 \gamma^2 (15\Omega^3) + \frac{1}{16}\gamma^4(105\Omega^4) \\
\mathrm{E}(\Delta V^5) &= \frac{5}{2}\delta^4\gamma(15\Omega^3) + \frac{5}{4}\delta^2\gamma^3(105\Omega^4)+ \frac{1}{32} \gamma^5 (945\Omega^5)
\end{align*}

\begin{figure}
\begin{subfigure}{.5\textwidth}
  \centering
  \includegraphics[width=.9\linewidth]{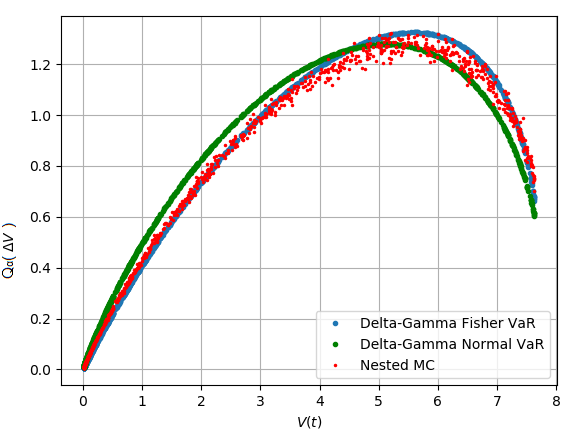}  
  \caption{Comparison of conditional quantiles at 1\% 
  for the call combination at a sample time $t$.}
  \label{comparevardeltagamma}
\end{subfigure}
\begin{subfigure}{.5\textwidth}
  \centering
  \includegraphics[width=.9\linewidth]{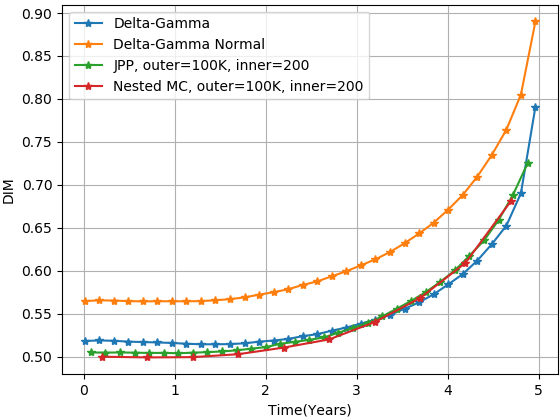}  
  \caption{Time Evolution of Dynamic Initial Margin}
  \label{comparedimdeltagamma}
\end{subfigure}
\caption{Comparison of Delta-Gamma approaches to estimate conditional quantiles
and time-evolution of DIM for the call combination.}
\end{figure}

The central moments, ($cm_1, cm_2, cm_3, cm_4, cm_5$) from the above raw
moments, the skew ($\kappa_3$), kurtosis ($\kappa_4$) and
$\kappa_5=cm_5^5/cm_2^{5/2}$ derived from above terms are then used to determine
the corresponding $\alpha$th quantile from Equation (\ref{cfexpansion}). The
comparison of DIM and VaR estimates obtained via this method for the
call combination discussed earlier is shown in Figure 
\ref{comparevardeltagamma} and Figure \ref{comparedimdeltagamma}, which is seen
to be in much closer agreement with benchmark Nested-Monte Carlo method, while
requiring far lower computational resources than the same. Hence the above
method is suitable for instruments and portfolios with given analytical or
approximate first and second order sensitivities.

\section{Nested Monte-Carlo with Moderate Inner Samples}\label{reducedinnersamples}
In the previous sections, the primary focus was on methods that only use outer
simulations to estimate properties of the distribution of $\Delta V$, such as
conditional quantile regression, quantile methods based on moment-matched
distributions and regressed moments, or methods that use sensitivities and
assumed distributions of  underlying risk factors returns. However, these
methods might benefit from refined estimates possible by limited  inner
simulations. The methods presented in this section below use the additional
information from small and limited number of inner simulations to further refine
their estimates for moment matching or quantile regression. Hence, the methods
in this section could be thought of as a possible middle-ground between
full-fledged Nested Monte-Carlo simulation with sufficiently many inner
simulations requiring large computational resources on one side and efficient
computational approaches such as moment-matching and quantile regression on the
other.

\subsection{Moment-Matching Methods}

\begin{figure}
    \centering
    \includegraphics[width=\linewidth]{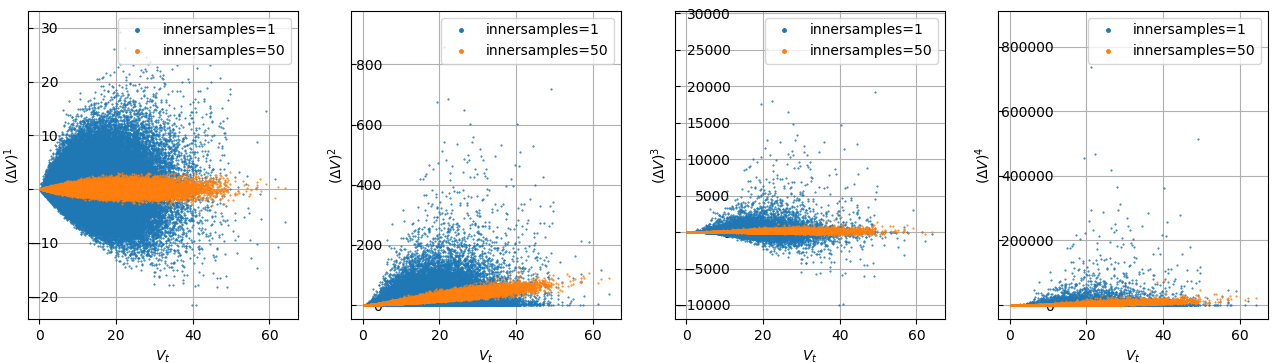}
    \caption{Refinement of first 4 raw moments at time 0.16 years with $N_I=50$ inner samples }
    \label{fxcallrawmomentswinnersamples}
\end{figure}

\begin{figure}
    \centering
    \includegraphics[width=0.6\linewidth]{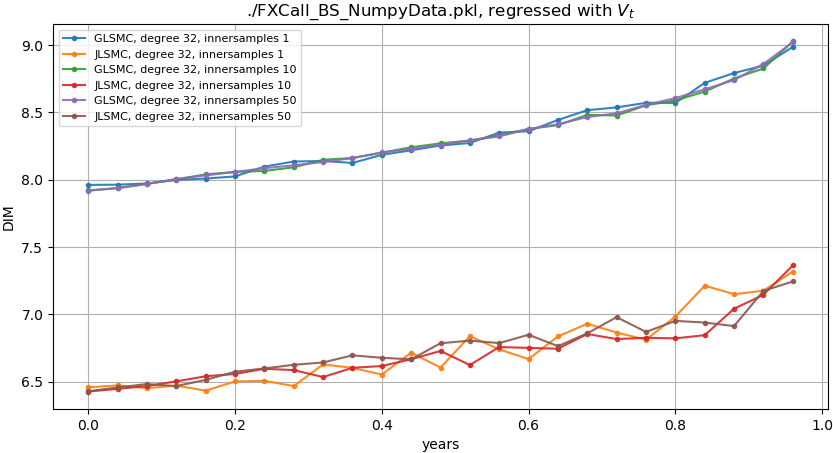}  
    \caption{Comparison of the time-evolution of DIM between the GLSMC and
    JLSMC with different number of inner samples.}
    \label{fxcalldiminnersamples}
\end{figure}

Based on the GLSMC and JLSMC methods presented in earlier sections, the
moment-matching algorithms were applied to moment estimates of $\Delta V|V$ that
used additional limited number of inner samples. The raw moments $(\Delta
V_m)^i$ for $i=1, 2, 3, 4$  from a single outer path are replaced by $(1/N_I)
\sum_{j=1}^{N_I} (\Delta V_{jm})^i$ for $i=1, 2, 3, 4$, where $N_I$ is the
number of inner simulations for the outer portfolio scenario $m$. Here the
number of inner simulations was chosen to be rather small as $N_I=50$ in order
to limit the needed computational resources. The comparison of raw moments is
shown in Figure \ref{fxcallrawmomentswinnersamples} at a specific time $t=0.16$
years, across all the outer paths. It shows that additional inner samples
decrease the variance of the raw moments. The time evolution of DIM obtained via
GLSMC and JLSMC methods using these raw moments is shown in Figure
\ref{fxcalldiminnersamples}. Although the time evolution obtained via GLSMC
method has not changed much, as it depends only on the first two moments, the
JLSMC method seems to exhibit a smoother time-evolution as it depends on and is
sensitive to higher moments.

\subsection{Johnson Percentile-Matching Method}
The Johnson Percentile matching method fits a Sl, Su or Sb type Johnson
distribution given a limited number of inner samples, based on a discriminant
computed from particular percentiles of the inner samples data. The procedure is
outlined in \cite{slifker1980, george2011}. In short, a fixed value $z (0< z<
1)$ is chosen and the corresponding percentiles $P_{3z},P_{-3z},P_z, P_{-z}$ of
the standard normal distribution corresponding to the four points $\pm 3z$ and
$\pm z$ are computed.
The values of data points corresponding to the percentiles are then determined
from the sample distribution as $Q_{P_i}(X)$, where $i=3z, z, -z, -3z$, as
$x_{3z}, x_z, x_{-z}, x_{-3z}$.

The value of the discriminant $d=mn/p^2$ is then used to select the type of
Johnson distribution, where $p=x_z-x_{-z}, m=x_{3z}-x_z, n=x_{-z}-x_{-3z}$. The
Su Johnson distribution is fit to the data if $d>1.001$, the Sb Johnson
distribution chosen if $d<0.999$ and the Sl Johnson distribution is chosen if
$0.999\leq d \leq 1.001$. This method provides closed form solution to the
Johnson distribution parameters $\xi, \gamma, \delta$ and $\lambda$ in terms of
the discriminant $d$ and $m, n$ hence requiring far lower computational
resources than the moment matching algorithm.

The time evolution of DIM obtained via this method with a much reduced $N_I=200$
number of inner simulations is compared with Nested Monte-Carlo and Delta-Gamma
approaches in Figure \ref{comparedimdeltagamma} and is seen to be in good
agreement with the Nested Monte-Carlo approach.

\subsection{Conditional Quantile Regression}
\begin{figure}
    \centering
    \includegraphics[width=0.6\linewidth]{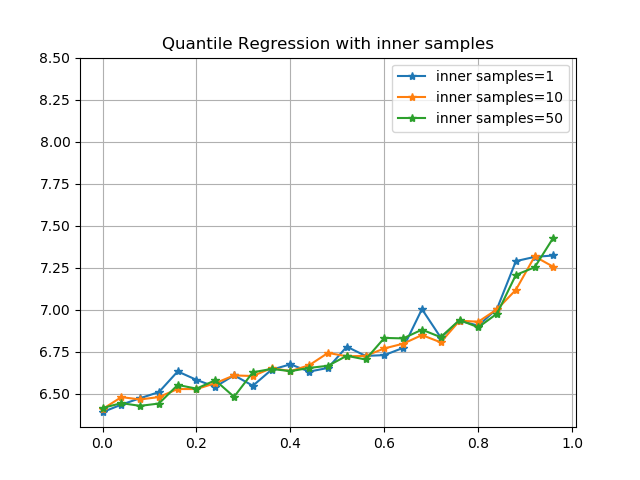}
    \caption{Comparison of the time-evolution of DIM via quantile regression 
    with different number of inner samples.}
    \label{quantilereginnersamples}
\end{figure}

The conditional quantile regression approach can also take advantage of
additional inner samples. Similar to the other methods, samples from limited
inner simulations were added to the training samples for quantile regression. A
comparison of the time-evolution of DIM obtained via quantile regression with
different number of inner samples is shown in Figure
\ref{quantilereginnersamples}. Although the inner samples help refine the
distribution, the time-evolution so obtained is still comparable in terms of
accuracy and scale to that obtained with a single inner sample. This indicates
that with sufficient number of outer samples (here $N_O=100,000$) the quantile
regression procedure can be robust with respect to variations in inner
distributions across the underlier at any given time or at different times.
Comparing the results with a single inner sample versus the others, it seems
that adding at least a few inner samples results in a somewhat smoother
evolution of the DIM.

\subsection{Pseudo-Inner Samples}
As the process of running nested inner simulations for a large number of outer
simulations is computationally very expensive, an alternative method
is to approximate inner simulations with available nearby outer simulation
values around a neighborhood of $X_t$ or $V_t$. Hence under some circumstances,
the values from outer simulations can be considered as proxy to samples from the
actual distribution of portfolio values or pseudo-inner samples.

\begin{figure}
\begin{subfigure}{.5\textwidth}
    \includegraphics[width=\linewidth]{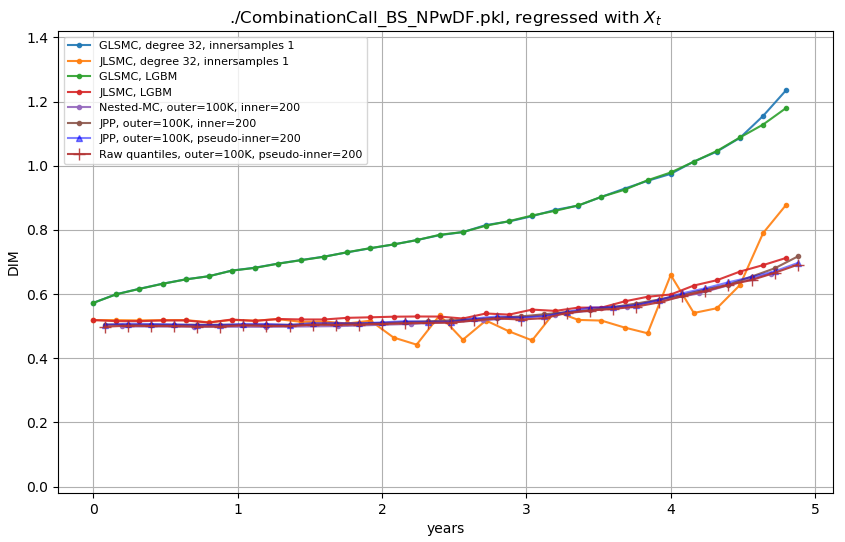}
    \caption{Callcombination with pseudoinner samples.}
    \label{combcalldimpseudosamples}
\end{subfigure}
\begin{subfigure}{.5\textwidth}
    \includegraphics[width=\linewidth]{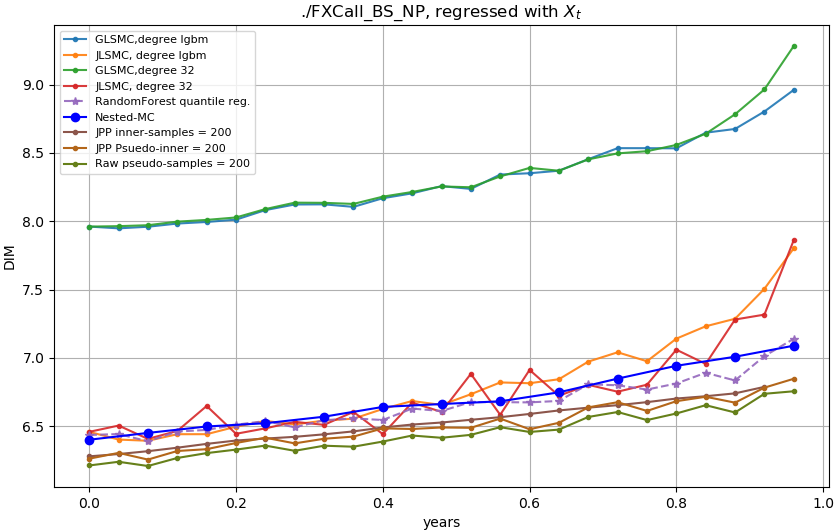}
    \caption{FXcall with pseudoinner samples.}
    \label{fxcalldimpseudosamples}
\end{subfigure}
\caption{Comparison of the time-evolution of DIM with pseudo-inner samples.}
\label{pseudosamples}
\end{figure}

The distribution $(\Delta V|x_t = X_t)$ is approximated by observations from
portfolio scenario $m$, $\Delta V_m$ given by equation \eqref{deltavpathm} such
that $x^{(m)}_t \in \mathcal{B}_\epsilon(X_t)$ or $V^{(m)}_t \in
\mathcal{B}_{\epsilon_1}(V_t)$ where the underlying asset or portfolio value is
observed to be within a neighborhood $\mathcal{B}_\epsilon$ of $X_t$ or $V_t$.
Figure \ref{pseudosamples} presents the time evolution of initial margin
estimates obtained via limited number of pseudo-inner samples considered around
around the neighborhood of $X_t$, for the Johnson's Percentile matching method,
along with regular inner-samples also implemented for the same method. It can be
seed that both JPP with inner samples and JPP with pseudo-inner samples agree
well with DIM obtained via Nested MC for the the combination call and FX call
instruments. Additionally, the DIM obtained by estimating quantiles from the raw
pseudo-samples is also presented, which is seen to be in good agreement with
that obtained via Nested MC as well.

\section{Comparing Running Times and RMSE}\label{methodcomparison}

We compare the methods presented above in terms of accuracy and speed. The
Nested MC method is considered to be the benchmark for estimating the initial
margin, and hence was used to measure the deviation in terms of RMSE accuracy.
Additionally, the raw run time of the various methods was also recorded in order
to compare multiple aspects of the methods. We ran the tests on a computer with
32GB of memory and Intel Core i7 processor with 8 cores.
We did not spend much time optimizing the run time of the different methods or
otherwise trying to ensure that the running times are very comparable so the
listed running times should be taken only as indicative.

Table \ref{fxcallres1} presents the results for the FX call option with DIM
computed for 25 times corresponding to various methods. 

In terms of running time for this example, nested Monte Carlo, Johnson Moment
Matching MC using LGBM regression and Johnson Percentile method take the
longest, with DNN quantile regression being faster than those but slower than
the quantile regressions with LGBM, LGBM quantile regressions take a medium
amount of time, and the other methods take less than 100s with some taking less
than 10s.

In terms of RMSE, the two quantile regressions using the LightGBM quantile
regression were the most accurate while all other methods except the GLSMC 
and Delta-Gamma Normal methods were in the same order of magnitude with 
respect to RMSE.

The JPP method was implemented also with pseudo-samples, where a suitable
Johnson distribution was fit to a collection from samples from adjacent outer
simulations taken in-lieu of actual inner simulation samples. Since the
percentile matching procedure has a convenient closed form solution, the
procedure can be vectorized for multiple outer simulation values corresponding
to different $V_t$s or $X_t$s, thus making for an efficient implementation for
VaR or IM estimation. The DIM results obtained via this procedure, without the
need for inner-samples, are compared against the previously studied methods. The
run times are significantly faster along with improved accuracy, compared to
traditional methods.

In addition, the same procedure can be repeated to estimate quantiles at each
$V_t$ or $X_t$ from raw (pseudo-inner) samples without the need to fit a Johnson
distribution. The results from this approach are also seen to outperform the
traditional approaches both in terms of run-time and accuracy.

Table \ref{ccallres1} presents results for a call combination with DIM computed
for 125 times corresponding to various methods. 

In terms of running time, we see similar results,  nested Monte Carlo, Johnson
Moment Matching MC using LGBM regression and Johnson Percentile method take the
longest, LGBM quantile regressions take a medium amount of time, and the other
methods take less than 100s with some taking less than 10s.

In terms of RMSE, using pseudo samples raw seems to give best accuracy,
followed by Johnson Percentile methods with pseudo samples followed by 
Johnson Percentile method, quantile regressions using LGBM, and Johson MC Moment
Matching using LGBM. On the other end of the spectrum, GLSMC methods, 
JLSMC regressing against $V_t$ (due to the non-uniqueness of the $V_t$ with
respect to $X_t$) and Delta-Gamma-Normal methods give the least accurate results
as measured by RMSE. 

We see that even across these two relatively simple examples for which we can
perform benchmarking with nested MC with large enough number of samples, a
rather large variety of methods performs reasonably well in terms of RMSE
accuracy and none performs uniformly best.

\begin{minipage}{\linewidth}
\centering
\footnotesize
\begin{tabular}{|r|c|c|c|}\hline
\thead{Method}&\thead{RMSE}&\thead{Time}&\thead{Specification} \\
\hline
\makecell{Nested MC} &   & \makecell{2375s$\pm$ 43s}   & 
\makecell{Outer=100K, Inner=2000} \\
\hline
\makecell{Raw Pseudo Samples} & 3.5e-5 & 16.2s$\pm$ 0.1s &
\makecell{Outer=100K, Pseudo-inner=200}  \\
\hline
\makecell{Johnson Percentile} &  1.25e-04 & \makecell{2568s$\pm$ 18s} & 
\makecell{Outer=10K,Inner=200} \\
\hline
\makecell{Johnson Percentile with Pseudosamples} & 8.3e-05 & 27.1s$\pm$ 0.3 &
\makecell{Outer=10K,Pseudo-inner=200, stride=10}\\
\hline
\makecell{Quantile Regression - LGBM($V_t$)}  & 3.93e-4 &  \makecell{269s$\pm$
12s} & \makecell{Outer=100K, LightGBM, Reg with $V_t$} \\
\hline
\makecell{Quantile Regression - LGBM($X_t$)}  & 6.96e-4 & 
\makecell{273s$\pm$ 13s} & \makecell{Outer=100K, LightGBM, Reg with $X_t$} \\
\hline
\makecell{Delta Gamma} & 7.33e-4 &  42.4s$\pm$ 0.9s & \makecell{Outer=100K} \\
\hline
\makecell{Delta Gamma Normal} & 7.7e-3 & 35.4s$\pm$ 0.7s &
\makecell{Outer=100K}\\
\hline
\makecell{JLSMC - LGBM($X_t$)} & 4.86e-4  & \makecell{2412s$\pm$ 45s} &
\makecell{Computed at 200 points in body and tails of $X_t$ \\ Outer=100K,
LightGBM, Reg with $X_t$} \\
\hline
\makecell{JLSMC  - Lag($X_t$)} & 6.8e-3 & 63.4s$\pm$ 2.3s& \makecell{Computed at
200 points in body and tails of $X_t$ \\ Outer=100K, Laguerre, order 32, Reg
with $X_t$} \\
\hline
\makecell{JLSMC - Lag($V_t$)} & 1.47e-2 & 58.3s$\pm$ 1.1s & \makecell{Computed
at 200 points in body and tails of $V_t$ \\ Outer=100K, Laguerre, order 32, Reg
with $V_t$} \\
\hline
\makecell{GLSMC - LGBM($X_t$)} & 8.94e-2 & 1220s$\pm$ 13s & \makecell{Computed at
200 points in body and tails of $X_t$,\\ Outer=100K, LightGBM, Reg
with $X_t$} \\
\hline
\makecell{GLSMC - Lag($X_t$)} & 9.7e-2 & 33.7s$\pm$ 1.3s & \makecell{Computed at
200 points in body and tails of $X_t$,\\ Outer=100K, Laguerre, order 32, Reg
with $X_t$} \\
\hline 
\makecell{GLSMC - Lag($V_t$)} &  8.9e-2 & 30.1s$\pm$ 0.9s & \makecell{Computed
at 200 points in body and tails of $V_t$,\\ Outer=100K, Laguerre, order 32, Reg
with $V_t$} \\
\hline
\end{tabular}
\captionof{table}{Comparison of performance metrics of various methods for the
Combination Call instrument \\with 125 DIM instances}\label{ccallres1}
\end{minipage}

\begin{minipage}{\linewidth}
\centering
\footnotesize
\begin{tabular}{|r|c|c|c|}\hline
\thead{Method}&\thead{RMSE}&\thead{Time}&\thead{Specification} \\
\hline
\makecell{Nested MC} &   & \makecell{438s$\pm$ 6s}   & 
\makecell{Outer=100K, Inner=2000} \\
\hline
\makecell{Raw Pseudo Samples} & 6.69e-2 & 3.28s$\pm$0.2s &
\makecell{Outer=100K, Pseudo-inner=200}  \\
\hline
\makecell{Johnson Percentile} &  2.74e-2 & \makecell{438s$\pm$ 9s} & 
\makecell{Outer=10K,Inner=200} \\
\hline
\makecell{Johnson Percentile with Pseudosamples} & 3.97e-02 & 5.97s$\pm$0.2 &
\makecell{Outer=10K,Pseudo-inner=200, stride=10}\\
\hline
\makecell{Quantile Regression - LGBM($V_t$)}  & 4.42e-3 &  \makecell{53s$\pm$
0.9s} & \makecell{Outer=100K, LightGBM, Reg with $V_t$} \\
\hline
\makecell{Quantile Regression - LGBM($X_t$)}  & 4.26e-3 & 
\makecell{52.6s$\pm$0.8s} & \makecell{Outer=100K, LightGBM, Reg with $X_t$} \\
\hline
\makecell{Quantile Regression - DNN}  & 4.7e-2 &  \makecell{281.6s$\pm$
15s} & \makecell{Outer=100K, Layers=4, Units=3, Batch-1024,\\ Minimization Steps=20K} 
\\
\hline
\makecell{Delta Gamma} & 3.38e-2 &  10.6s$\pm$ 0.8s & \makecell{Outer=100K} \\
\hline
\makecell{Delta Gamma Normal} & 1.72 & 9.5s$\pm$ 0.8s &
\makecell{Outer=100K}\\
\hline
\makecell{JLSMC - Lag($V_t$)} & 1.11e-2 & 12.82s$\pm$ 1.5s & \makecell{Computed
at 200 points in body and tails of $V_t$ \\ Outer=100K, Laguerre, order 32, Reg
with $V_t$} \\
\hline
\makecell{JLSMC  - Lag($X_t$)} & 3.82e-2 & 13.4s$\pm$1.3s& \makecell{Computed at
200 points in body and tails of $X_t$ \\ Outer=100K, Laguerre, order 32, Reg
with $X_t$} \\
\hline
\makecell{JLSMC - LGBM($X_t$)} & 4.13e-2 & \makecell{487s$\pm$7s} &
\makecell{Computed at 200 points in body and tails of $X_t$ \\ Outer=100K,
LightGBM, Reg with $X_t$} \\
\hline
\makecell{GLSMC - Lag($V_t$)} &  2.54 & 7.4s$\pm$ 1.2s & \makecell{Computed
at 200 points in body and tails of $V_t$,\\ Outer=100K, Laguerre, order 32, Reg
with $V_t$} \\
\hline
\makecell{GLSMC - Lag($X_t$)} & 2.67  &  8.1s$\pm$0.9s & \makecell{Computed at
200 points in body and tails of $X_t$,\\ Outer=100K, Laguerre, order 32, Reg
with $X_t$} \\
\hline 
\makecell{GLSMC - LGBM($X_t$)} & 2.57 & 245s$\pm$4s & \makecell{Computed at
200 points in body and tails of $X_t$,\\ Outer=100K, LightGBM, Reg
with $X_t$} \\
\hline
\end{tabular}
\captionof{table}{Comparison of performance metrics of various methods for the
FX Call instrument \\with 25 DIM
instances}\label{fxcallres1}
\end{minipage}

\subsection{Discussion and Guidelines}
It is illustrative to reflect on the requirements and constraints of each of 
these methods in light of the above results. The methods with pseudo samples 
outperform the traditional JLSMC and Quantile Regression methods in terms of 
accuracy and speed for these particular financial instruments with 1-dimensional 
risk-factors ($X_t$) and portfolio values ($V_t$). However, with multiple risk 
factors in higher dimensions the number of pseudo samples that can be borrowed 
from within a hyper-bin would be much sparser, thus leading to noisy estimates 
of quantiles. Secondly the number of hyper-bins that need to be considered grows 
rapidly with the dimensions thus forcing one to resort to $V_t$ based 
pseudo-samples for quantile estimation. Furthermore, even for 1-dimensional 
problems, it is necessary to have sufficient number of samples within a 
neighborhood of $V_t$ or $X_t$, in order to accurately capture the properties of 
the distribution. The number of active samples within any neighborhood of 
$V_t$ or $X_t$ becomes sparse with higher volatility and time horizon, again 
leading to noisy estimates. Hence, the pseudo-sample based methods are ideally 
suited for problems with rich outer-simulation data.

Methods with moment regression (JLSMC) give medium accuracy wherein it is 
necessary to have a good estimate of moment values, with methods and
distributions that can fit higher moments typically performing better (JLSMC
performing better than GLSMC). 

Any inconsistencies in moment estimation leads to undefined distributions 
thus necessitating moment correction. Hence moment regression methods are 
particularly sensitive to the type of moment regression and the corresponding 
moment matching procedure. We also see that moment regression methods are 
relatively sensitive to the exact way how they are set up, thresholds, etc. 

The quantile regression (LGBM quantile regression and Deep Quantile regression)
procedures perform best with sufficient number of outer samples as they aim to
minimize the quantile regression loss function. However the computational
requirements grow with the number of available samples (outer-paths) while
requiring sophisticated optimization frameworks. It is seen that LGBM quantile
regression produces results comparable to pseudo-sample methods, as it employs a
series of tree based optimizations and empirical estimation of quantiles in
order to minimize the quantile loss function at the cost of longer runtime. On
the other hand, the Deep Quantile regression procedure involves minimizing the
quantile loss function via neural networks over several thousand minimization
steps, which can become impractical for large number of DIM steps, unless the
training of and optimization over the deep neural networks can be sped up
substantially.

The objective of this study is to present the details and procedures of various
methods used for estimation of fVaR and initial margin, along with with an
objective evaluation of their accuracies and performance. The choice of method
for a particular problem or financial instrument depends on variety of
constraints and requirements as outlined in Table \ref{reqmatrix} and there is
probably not a single method that can be applied for all financial instruments
under all scenarios.

\section{Conclusion}\label{conclusion}

This paper discussed several methods to estimate future Values at Risk or margin
requirements and their expected time-evolution (also called Dynamic Initial
Margin), from simple options to more complex IR swaps. The methods analyzed in
this paper are suited for instruments whose portfolio values are known or can be
approximated well pathwise. In order to establish a baseline, the Nested MC
method which is traditionally considered as a benchmark was used to validate
other classical methods to estimate DIM. Under the current assumptions, the
accuracy of these methods were studied taking into account necessary
computational resources (Nested MC, moment matching, quantile regression,
Johnson's percentile-matching) and when first and second order sensitivities are
available (Delta-Gamma methods).
Furthermore, the differences between the estimated Dynamic Initial Margin curves
were highlighted between the simpler GLSMC (Gaussian Least Square Monte-Carlo)
and Delta-Gamma Normal and their more complex counterparts, JLSMC (Johnson Least
Square Monte-Carlo) and Delta-Gamma methods. It was shown that standard moment
regression methods can lead to violations of the skew-kurtosis relationship and
that a simple correction procedure can lead to valid Johnson distributions and
better performance. Conditional quantile regression with or without deep
learning is a suitable technique with the help of an optimization framework,
which however can demand computational resources for large number of DIM steps.
The above methods were also compared in the presence of limited number of inner
simulation runs to see whether such inner samples would further improve or
refine the conditional quantile estimates.
While all the methods benefit from available inner samples, it was seen that
quantile regression tends to be more robust with sufficiently large number of
outer simulation runs, even without additional inner samples. Finally the
techniques where inner samples are necessary, such as Nested MC and Johnson's
Percentile matching, it is possible to replace them with pseudo-inner samples
under specific conditions, for comparable performance and much faster runtimes
where the results were demonstrated for FX call and combination call
instruments.

\section{Acknowledgement}
The authors would like to thank Vijayan Nair for valuable discussions and
comments regarding methods (in particular his suggestions to explore
pseudo-sample approaches), presentation, and results, and Agus Sudjianto for
supporting this research.

\bibliographystyle{plain}
\bibliography{./dimrefs}

\begin{thebibliography}{10}

\bibitem{albanese2020xva}
Claudio Albanese, Stephane Crepey, Rodney Hoskinson, and Bouazza Saadeddine.
\newblock {XVA} analysis from the balance sheet.
\newblock {\em arXiv preprint arXiv:2009.00368}, 2020.

\bibitem{book:alexander08}
Carol Alexander.
\newblock {\em Market Risk Analysis. Volume IV: Value-at-Risk Models}.
\newblock John Wiley \& Sons, Ltd, 2008.

\bibitem{caspers2017}
Peter Caspers, Paul Giltinan, Roland Lichters, and Nikolai Nowaczyk.
\newblock Forecasting initial margin requirements - a model evaluation.
\newblock 2017.
\newblock Available at SSRN: \url{https://ssrn.com/abstract=2911167} or
  \url{http://dx.doi.org/10.2139/ssrn.2911167}.

\bibitem{george2011}
Florence George and K.~M. Ramachandran.
\newblock Estimation of parameters of {J}ohnson’s system of distributions.
\newblock {\em Journal of Modern Applied Statistical Methods}, 10(2):494--504,
  2011.

\bibitem{hill1976}
I.~Hill, R.~Hill, and R.~Holder.
\newblock Algorithm {AS 99}: Fitting {J}ohnson curves by moments.
\newblock {\em Journal of the royal statistical society. Series C (Applied
  statistics)}, 25(2):180--189, 1976.

\bibitem{johnson1949}
N.~L. Johnson.
\newblock Systems of frequency curves generated by methods of translation.
\newblock {\em Biometrika}, 36(1/2):149--176, 1949.

\bibitem{jones2014}
D.~Jones.
\newblock {J}ohnson curve toolbox for {M}atlab: analysis of non-normal data
  using the {J}ohnson family of distributions.
\newblock 2014.

\bibitem{koenker2001}
Roger Koenker and Kevin~F. Hallock.
\newblock Quantile regression.
\newblock {\em Journal of Economic Perspectives}, 15(4):143--156, 2001.

\bibitem{mcwalter2018}
Thomas McWalter, Joerg Kienitz, Nikolai Nowaczyk, Ralph Rudd, and Sarp~Kaya
  Acar.
\newblock {D}ynamic {I}nitial {M}argin estimation based on quantiles of
  {J}ohnson distributions.
\newblock 2018.
\newblock Also Available at SSRN: \url{https://ssrn.com/abstract=3147811} or
  \url{http://dx.doi.org/10.2139/ssrn.3147811}.

\bibitem{book:shreve04}
Steven Shreve.
\newblock {\em Stochastic Calculus for Finance II, Continuous-Time Models}.
\newblock Springer-Verlag New York, 2004.

\bibitem{slifker1980}
James~F. Slifker and Samuel~S. Shapiro.
\newblock The {J}ohnson system: {S}election and parameter estimation.
\newblock {\em Technometrics}, 22(2):239--246, 1980.

\end{thebibliography}

\end{document}